\documentclass[%
 reprint,
 amsmath,amssymb,
 aps,
]{revtex4-2}

\usepackage{graphicx}
\usepackage{dcolumn}
\usepackage{bm}
\usepackage{hyperref}
\usepackage{color}

\begin{document}

\preprint{APS/123-QED}

\title{Violation of triboelectric charge conservation on colliding particles}

\author{Felix Jungmann}
 \altaffiliation[felix.jungmann@uni-due.de ]{}
\author{Hannah van Unen}%
\author{Jens Teiser}%
\author{Gerhard Wurm}%
\affiliation{%
 University of Duisburg-Essen, Lotharstr. 1, D-47057 Duisburg, Germany 
}%

\date{\today}

\begin{abstract}
In microgravity experiments, we quantified the net charge on systems of two identical, 434 $\rm \mu m$ diameter glass spheres before and after a collision. 
We find that charge conservation is significantly violated. Independent of the sign of the total charge, the systems regularly lose some of their net charges, that is, they slightly discharge. This implies that positive as well as negative charge carriers become entrained into the surrounding atmosphere during a collision. 
\end{abstract}

\maketitle

\section{Introduction} 

When two sand sized-grains collide they charge \citep{Harper1967, Lacks2019}. As simple as this sounds, as manifold are the explanations, ranging from electron transfer over the role of adsorbates like water to material transfer \citep{Wiles2004, Mccarty2007, Duff2008, Waitukaitis2014, Lee2018, Pan2019, Grosjean2020, Mizzi2019}. So, there is currently still a debate going on, which effects are relevant in which setting. We would like to step back here and take a somewhat different point of view on this problem. 
If only the charge on the two particles involved mattered, the total net charge should be conserved whatever the details of the charge transfer mechanism on the surface might be. But is this the case?

\section{Experiments}

\begin{figure}
	\includegraphics[width=\columnwidth]{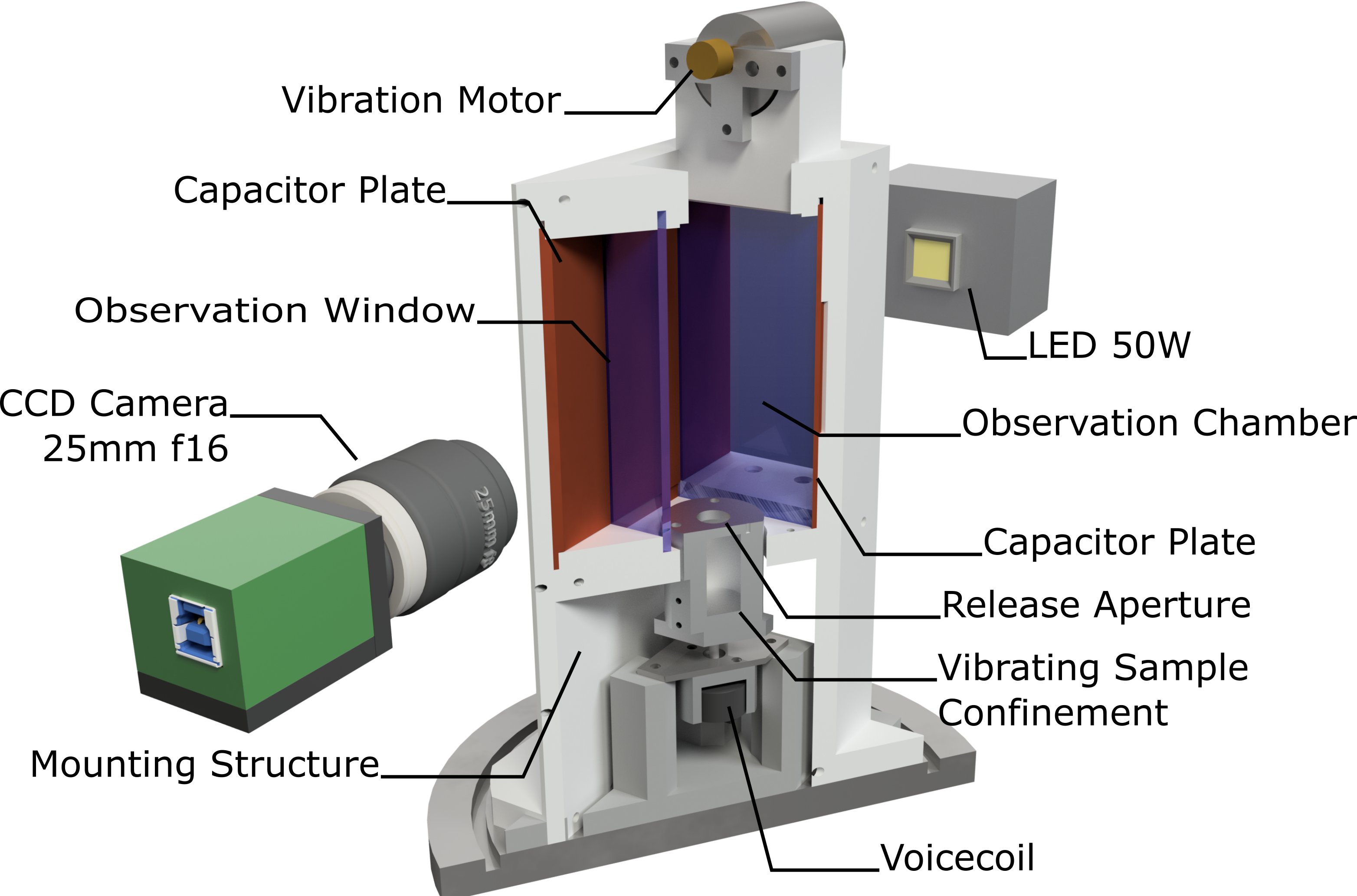}
	\caption{\label{fig:setup} Setup of the experiment taken from \citet{Steinpilz2020a}.}
\end{figure}

One way to measure the small charges on two grains before and after a collision is to analyze the four particle trajectories within an electric field under microgravity, provided by the drop tower in Bremen (Germany). Here, slow grains can be observed in detail and undisturbed otherwise. A sketch of the experiment can be seen in Fig. \ref{fig:setup}. This is the same as used in \citet{Steinpilz2020a}. Glass grains with a diameter of $434 \pm 17\,\rm \mu m$ and a mass of 0.1 mg were shaken in a cylinder coated with the same grains for 15 min, using a voicecoil. Due to the collisions they charge \citep{Steinpilz2020a, Steinpilz2020b, Jungmann2018, Jungmann2021}. When the 9\, seconds of microgravity begin, they are injected into a volume of $90 \times 48 \times 36\, \rm mm^3$ in size. The atmosphere in the volume is $\rm CO_2$ under normal pressure which is held quiescent. Particles are observed with a camera and bright field illumination is provided from the back. Perpendicular to the camera's perspective, copper capacitor plates are placed. Therefore, accelerations caused by the field are also aligned perpendicular to the camera's perspective. During microgravity, the residual gravity is smaller than $10^{-5}\, \rm g$, which makes the Coulomb force dominant when an electric field is applied. The strength of the field varied in different launches from 21 to 84 kV/m.
Suitable collisions (no close third grains, sufficient lengths of tracks before and after) are detected by eye and their trajectories tracked manually using Image J \citep{Rasband1997}. The charge of a grain $q$ is related to the accelerations $a$ on each trajectory by $q=m\,a/E$ with the electric field $E$ and the mass $m$ of the particle. The latter two are well-known in the experiments we carried out. 

\begin{figure}
	\includegraphics[width=\columnwidth]{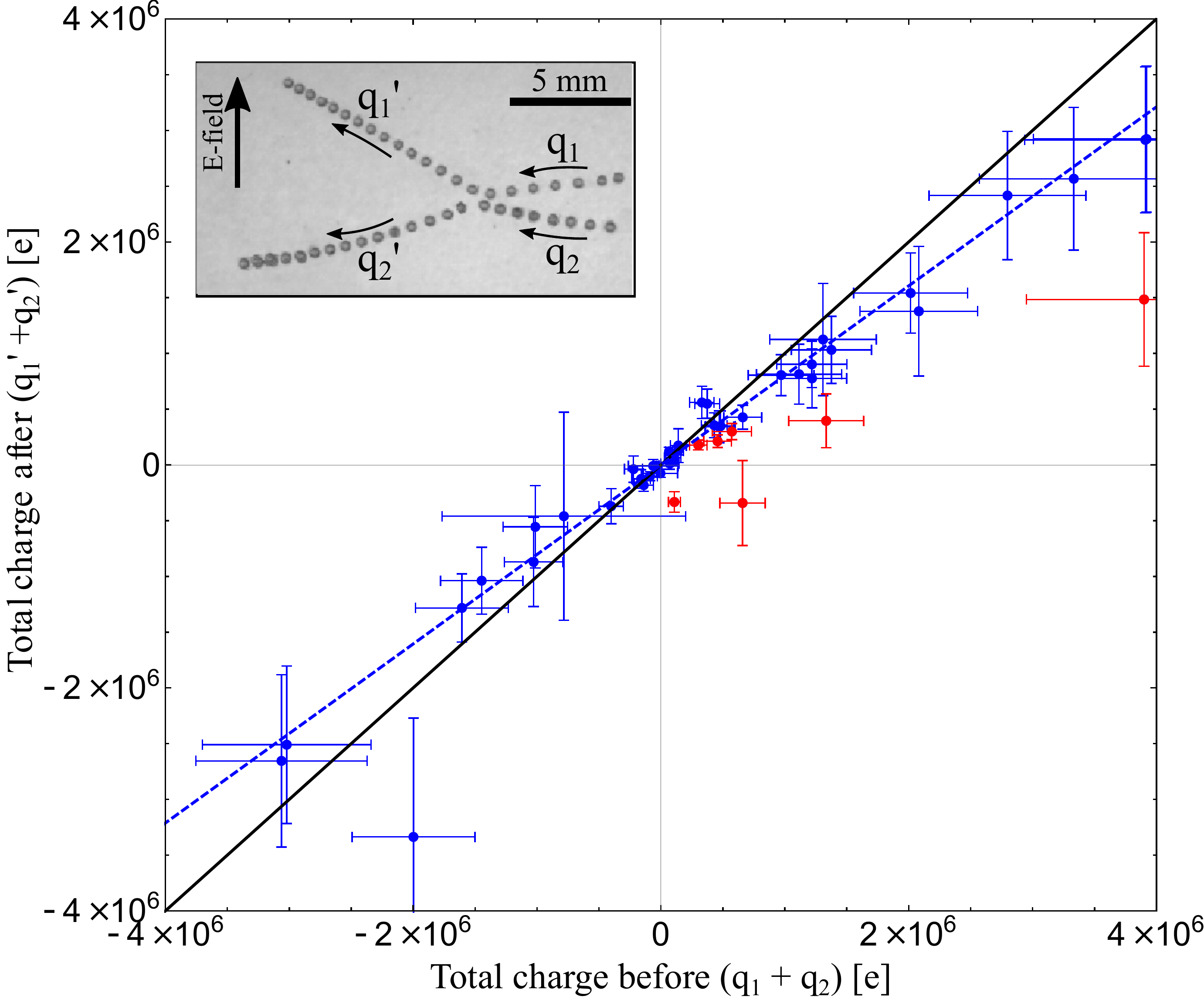}
	\caption{\label{fig:conservation} Total charge on two particles after ($q_1' + q_2'$) and before ($q_1 + q_2$) a collision. The bold black line marks the charge conservation (slope = 1). The dashed blue line is a linear trend to the data (slope = $0.80 \pm 0.03$).  Red square symbols mark cases of unambiguous violation of charge conservation. The inset in the top left is an example taken from the experiments to visualize the situation. Here, an overlay of 22 images during a collision is shown, covering a period of 120 ms. Other particles in the surroundings are removed.}
\end{figure}

Due to the homogeneous electric field of a plate capacitor the acceleration is constant and a track can be fitted by a parabola. Inter-particle attractions or repulsions are neglected as they are only dominating over the capacitor field very close to the actual collision contact \citep{Jungmann&Wurm2021}. Gas drag is notable for long tracks but not included explicitly for fitting. A typical velocity at the moment of the collision is 5 cm/s. Using this and Stokes drag one gets a typical maximum acceleration of $0.037\,\rm m/s^2$, which is, compared to the Coulomb acceleration ($0.267\,\rm m/s^2$), about 13\%. We added 13\% uncertainty of the net charges to the error bars in Fig. \ref{fig:conservation}.

This principle of charge measurements has been used in several experiments with similar setups \citep{Jungmann2018, Steinpilz2020a, Steinpilz2020b, Jungmann&Wurm2021}. We, therefore, refer the reader to these works for more details on the experiments. So far, we have never considered the collisions of two particles in more detail though. 

We do not try to investigate on which parameters the exact charge transfer within the collisions depends, as this requires a detailed analysis which we leave to future work.
We only consider the total net charge of both particles before and after a collision in Fig. \ref{fig:conservation}. A sketch of the situation is also included in Fig. \ref{fig:conservation}. 

\begin{figure}
	\includegraphics[width=\columnwidth]{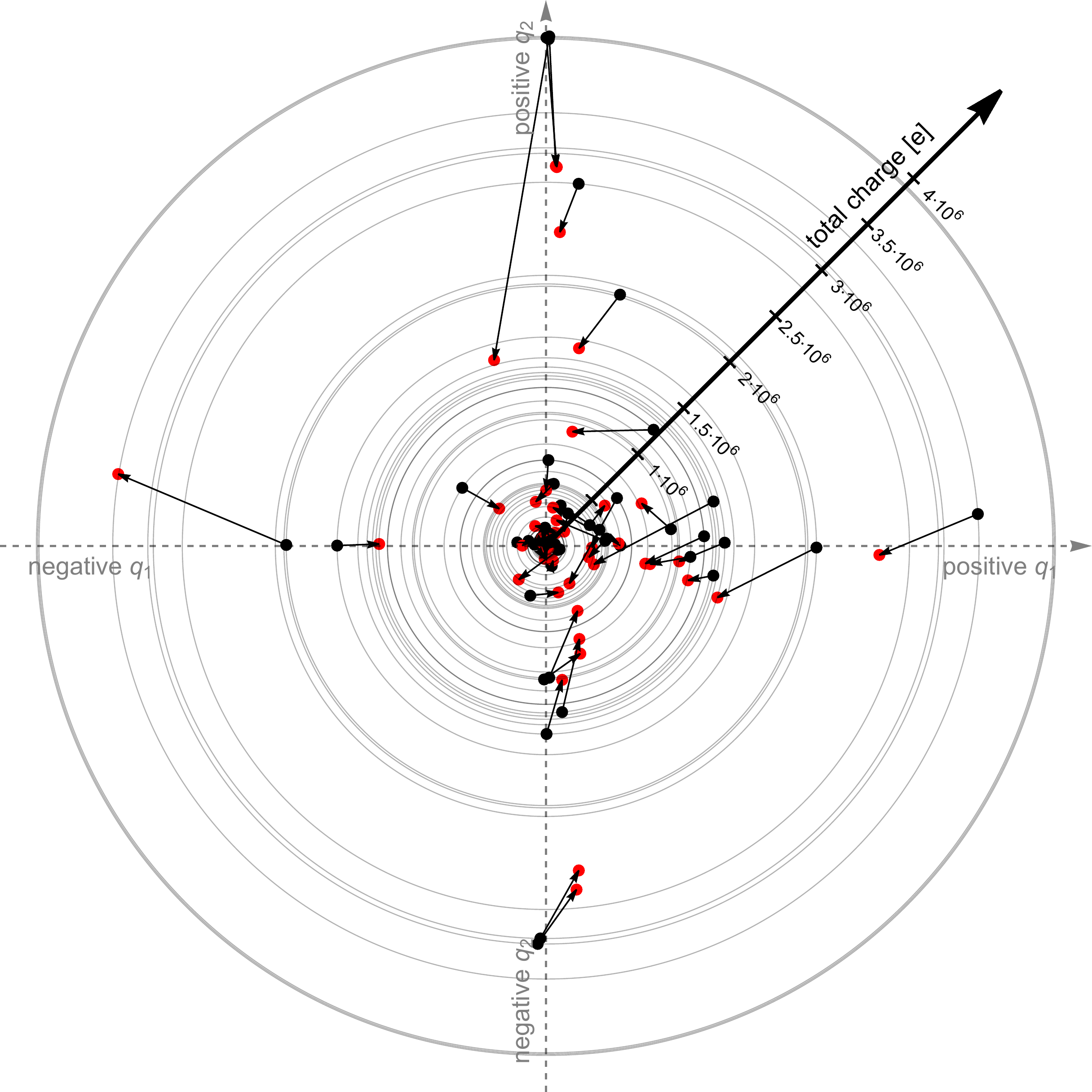}
	
	\includegraphics[width=\columnwidth]{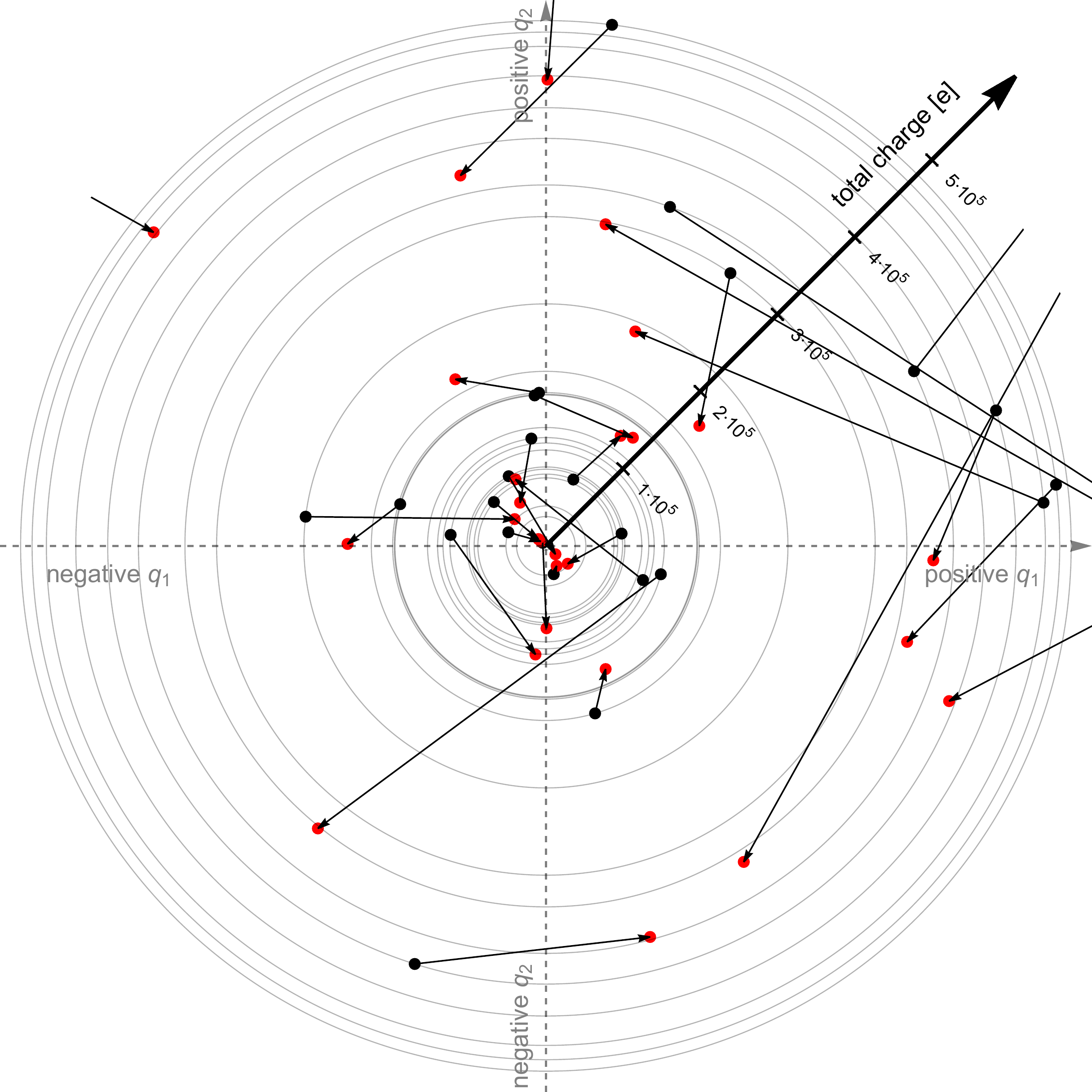}
	\caption{\label{fig:circles}Charge balance for different charge combinations in collisions between two particles. The radial distance of each data point to the center shows the sum of both charges ($q_1 + q_2$). Each quadrant marks a certain polarity combination. Due to the arbitrary choice of the first and second particle, the top left and bottom right quadrant are equivalent. The direction of the position vector corresponds to the specific charge ratio ($\phi = \arctan(q_1/q_2)$). Black and red dots mark the total charges before and after a collision, respectively. Inward directed arrows mark a loss in total charge, outward directed arrows mark a gain in total charge. Top: All measured values; Bottom: Zoom into the inner rings up to total charges of $5\cdot 10^5\,$e.}
\end{figure}

\section{Discussion}

Fig. \ref{fig:conservation} shows 
the total net charge of the two collision partners after the collision, plotted over the total net charge of the spheres before the collision. 
On average, on the level of a single particle, each particle's charge changes by $44 \pm 6 \%$, {that is in total numbers about $2 \cdot 10^{5}\, \rm e$. This is consistent with \citet{Kline2020}, who could measure similar charge transfer depending on the particle's material}. Uncertainties are due to the trajectory fitting, the uncertainty in mass, the uncertainty of the camera's perspective and the decelerating gas drag.
The solid line in Fig. \ref{fig:conservation} marks the scenario of charge conservation. Within their error bars, most individual data points are in agreement with charge conservation as it has also been argued by \citet{Kline2020} who studied two particle collisions in an acoustic trap. Especially for smaller charges, this appears to be the case. However, considering all the data, there seems to be a trend that the absolute net charges after a collision are smaller than before. Besides this subtle trend, there are several cases, where charge conservation is way beyond the error bar. These are no spurious outliers (Fig. \ref{fig:conservation}) but proof that the charge in collisions is just not regularly conserved on both grains.

Fig. \ref{fig:circles} shows the change in total charge in more detail in a polar plot. The directions of the position vectors correspond to the ratio between the involved charges ($\phi = \arctan(q_1/q_2)$). The radial distance of the data points to the center is scaled differently, so it shows the total charge ($q_1 + q_2$).

We note that we use identical grains, so the direction of charge transfer on weakly charged grains is random and it cannot be deduced from the bottom diagram with low charged grains how grains become highly charged. The surface is likely consisting of charge spots and a collision just adds another charge spot of arbitrary polarity \cite{Steinpilz2020b}. High net charges then require several collisions but this cannot be studied in our two particle collisions here.

Discharge is independent of polarity as most of the arrows point to the center (decreasing in net charge). Data-points are not equally distributed in the diagrams, i.e. collisions of two highly but oppositely charged grains are suppressed. This is an experimental bias due to the reduced collision probabilities of these cases. Both types of charges are attracted by different electrodes but as highly charged grains are more strongly accelerated they are more likely found closer to their respective electrode. In contrast, the initial momentum can carry less charged grains way closer to the repelling electrode enabling collisions between opposite-charge grains. In addition, two particles with very high velocities cannot be tracked by our method.

The behavior of less charged grains (zoomed-in plot) is a bit more chaotic due to higher charge transfers compared to the net charge, but in general, there is the same charge reduction independent of the magnitude of the total charge or polarities.

\subsection{Discharge options}

 The first reason why we observe less charge after a collision might be a slow but continuous discharge by atmospheric ions which are always present. As the grain motion is analyzed after a collision, we might observe the integrated loss over a period of up to a second, which would not be connected to the collision then.
This is not the case, since the tracks in the electric field match the motion induced by a constant acceleration in combination with gas drag well. This already rules out a continuous charge loss during this time of free-flight. 

To be more specific, we can also estimate the discharge expected by atmospheric ions. There are various sources of ionization in the atmosphere, cosmic radiation being important in a closed chamber as ours. In an earlier work \citet{Jungmann2021} studied its effect on collisional charging in stratospheric balloon experiments, finding it to be negligible on the timescales of seconds. In any case, a worst-case estimate for regular discharge is as follows:

The ion density in equilibrium at ground level 
is well below 1000 charges /$\rm{cm^3}$. Our chamber has less than 1000 $\rm{cm^3} $ with more than 10.000 grains. If atmospheric charges were present and if all free charges were collected by the grains instantly, each grain charge would only change by 100 elementary charges. We consider this instant discharge appropriate as recharging by cosmic radiation would require much longer timescales than given by the experiment \citep{Jungmann2021}. Any charge reduction is, therefore, orders of magnitude lower than the absolute grain charges measured. This estimate does not yet account for the fact that charges could already be collected prior to collisions and it also does not yet account for the fact that the capacitor might preferentially remove the charges first. Thus, overall, it is obvious that discharges can only occur during the collisions in a much faster process and regular atmospheric discharge can be neglected.

Certainly, if the charge is not conserved on the grains it has to go somewhere. As the experiments were carried out in a gaseous environment, ions and/or electrons obviously are set free into the atmosphere. This is not necessarily surprising. Atmospheric electricity builds on the ions from discharges of tribocharged grains during large scale thunderstorms \citep{Latham1981,Saunders2008,Saunders2008b,Pahtz2010}. If this also works on small scales \citep{Mccarty2007,Wurm2019,Harper2020,Schoenau2021}, charge might leak out here during atmospheric breakdown. Not only large net charges can cause an atmospheric breakdown. Even though we chose particles as smooth as possible on their surface, small asperities might be present. Quite clear, as electric fields surrounding sharp tips can be significantly higher, local breakdown might occur more frequently. In this context, also flexoelectricity might be important. High electric fields might be produced by material tension during collisions and this has only recently been suggested as another option for tribocharging \citep{Mizzi2019}. Especially, for rough surfaces, this might cause local ionization by an atmospheric breakdown. In any case, charges produced during atmospheric breakdown might easily diffuse into the gas and leave the two particle system.

Another possibility is that discharge might be due to the direct liberation of ions from the grain surfaces. Here, the transfer of water ions would be one favourite as it is already proposed for charging itself \citep{Kudin2008, Lee2018}. Even if experiments are carried out in a dry CO$_2$-atmosphere, molecular layers of water will be adsorbed on the surface of the glass spheres. 
Naturally, due to Coulomb repulsion and attraction, independent of the origin of ions, this will predominantly lead to the net discharge of the two particle system, as observed.

\section{Conclusion}

The environment plays an important role in the total charge balance during the collisions of grains.
The experiments presented here show that ions or charge carriers of both polarities somehow become entrained into the gas phase. As consequence, the total net charge on two particles in a collision is not conserved. This might be of relevance for the ionization fraction of an atmosphere and charge limits on grains. 
In any case, this short note highlights the fact that there is more
to charge transfer in a collision between two grains than the charge on the two grains themselves.

\acknowledgments

This project is supported by DLR Space Administration with funds provided by the Federal Ministry for Economic Affairs and Energy (BMWi) under grant numbers DLR 50WM1762 and 50WM2142. We appreciate the very constructive reviews of two anonymous referees. 

\def\aap{{Astron. and Astrophys.}}
\def\aapr{{Astron. and Asrophys. Rev.}}
\def\araa{{Annu. Rev. Astro. Astrophys}}
\def\aaps{{Astron. and Asrophys. Suppl.}}
\def\adsr{{Adv. Space Research}}
\def\aj{{Astronomical J.}}
\def\apj{{Astrophysical J.}}
\def\apjs{{Astrophysical J. Suppl.}}
\def\apjl{{Astrophysical J. Letters}}
\def\aspc{{Astronomical Soc. Pacific Conf. Ser.}}
\def\ass{{Astrophys. Space Sci.}}
\def\bala{{Baltic Astronomy}}
\def\cmp{{Contr. Mineralogy and Petrology}}
\def\cpc{{Computer Physics Comm.}}
\def\epsl{{Earth and Planetary Sci. Lett.}}
\def\gca{{Geochimica and Cosmochimica Acta}}
\def\ica{{Icarus}}
\def\jp{{J. Petrol.}}
\def\jqsrt{{JQSRT}}
\def\mn{{Monthly Notices Roy. Astr. Soc.}}
\def\mps{{Meteoritics and Planetary Sci.}}
\def\nat{{Nature}}
\def\phre{{Physical Rev. E}}
\def\pre{{Physical Rev. E}}
\def\prl{{Physical Rev. Lett.}}
\def\pss{{Planetary and Space Sci.}}
\def\rmf{{Rev.Mod.Phys}}
\def\sci{{Science}}
\def\ssr{{Space Science Rev.}}
\def\mnras{{Monthly Notices Roy. Astr. Soc.}}
\def\jgr{{J. Geophys. Res.}}
\def\icarus{{Icarus}}
\def\planss{{Planetary and Space Science}}
\def\grl{{Geophysical Research Letters}}
\def\apss{{Astrophysics and Space Science}}

\bibliography{references}{}

\end{document}